\documentstyle[twocolumn,aps,epsf]{revtex}
\begin{document}
\newcommand{\beq}{\begin{equation}}
\newcommand{\eeq}{\end{equation}}

\title{FOLDING IN TWO-DIMENSIONAL OFF-LATTICE MODELS OF PROTEINS}

\author{Mai Suan Li and Marek Cieplak}

\address{Institute of Physics, Polish Academy of Sciences,
Al. Lotnikow 32/46, 02-668 Warsaw, Poland }

\address{
\centering{
\medskip\em
{}~\\
\begin{minipage}{14cm}
Model off-lattice sequences in two dimensions are constructed
so that their native states are close to an on-lattice target.
The Hamiltonian involves the Lennard-Jones and harmonic interactions.
The native states of these sequences are determined
with a high degree of certainty
through Monte Carlo processes.
The sequences are characterized thermodynamically
and kinetically.
It is shown that the rank-ordering-based scheme of the assignment
of contact energies typically fails in off-lattice models even though it
generates high stability of on-lattice sequences.
Similar to the on-lattice case, Go-like modeling, in which the
interaction potentials are restricted
to the native contacts in a target shape, gives rise to
good folding properties. Involving other contacts deteriorates these
properties.
{}~\\
{}~
\end{minipage}
}}

\maketitle


\section{Introduction}
Understanding of the folding process of proteins is one
of the current challenges in molecular biophysics.
Many insights into the nature of folding have been provided by studying
lattice models in which a protein is
represented by a chain of beads on a hupercubic lattice 
(see for instance ref.\cite{Dill}).
A more realistic modeling of proteins, however, requires considering
off-lattice systems.
Simple off-lattice systems have been
discussed recently by Irback {\em et al}.\cite{Irback}
and Klimov and Thirumalai\cite{Klimov}. The former authors 
have studied a model with two kinds of residues
and they have found that very few such
sequences give rise to a rapid folding.

In this paper we focus on  ways to design off-lattice
sequences that form good folders.
Specifically, we consider a two-dimensional (2D) target shape and assign two 
models of interaction energies between the beads. The first model is
a generalization of the Go-like approach\cite{Go} to off-lattice situations. 
The second model, on the other hand, is a generalization of the 
rank-ordering-based assignment of the couplings\cite{Cieplak,Cieplak1}.
When the beads are constrained to be located on sites of a lattice, both models
provide sequences which are good folders. Here, we demonstrate that
this may not be true for off-lattice models. Namely, an
extended character of the interactions, that is necessarily present
in off-lattice Hamiltonians, gives rise to differing levels of frustration
in the two models and
leaves only the Go-like sequences as good folders.
This is compatible with the principle of minimum frustration proposed
by Bryngelson and Wolynes\cite{Bryngelson}.

We formulate our models in the context of the Hamiltonian used by
Iori, Marinari, and 
Parisi\cite{IMP} (IMP) in which
monomers interact via the Lennard-Jones potential. The 
amplitudes of the attractive part between beads $i$ and $j$
 of this potential, $A_{ij}$, 
representing a residue-dependent interaction, are quenched random variables. 
Additionally, the monomers are tethered
sequentially along the chain by means of harmonic interactions.
IMP and Struglia\cite{Struglia} have
demonstrated the existence of a compact phase in 
a 3D version of the model.
In these studies the dynamics have been defined in terms of a 
Monte Carlo process. The true ground state - the native state-- 
however, typically is not known (as in IMP) or there is a substantial
degree of uncertainty whether the state assumed to be native is
indeed the ground state (as in the paper by Struglia \cite{Struglia}).

Studies of the dynamics of folding require knowing the precise
shape of the native state conformation  without which the folding
time cannot be defined. For small scale lattice models,
the ground state may be obtained by an exact enumeration of conformations.
This method, however, does not apply to off-lattice models.
Here we present 2D Lennard-Jones sequences of 16 monomers in which
the ground state is known accurately and with
a high degree of certainty. We then
provide some basic characterization of these sequences, obtained through
the Monte Carlo procedure.

The construction of the model sequences is presented in Sec. II. We consider
the lattice target shown in Fig. 1(a). In order to generate a Go-like sequence,
$G$, we assign the $A_{ij}$ of the Lennard-Jones potential to be uniform
in the native contacts and zero in the non-native contacts, i.e., all pairs
of monomers that do not form contacts in the native state interact only 
through a short-range repulsion. The second model sequence, $R'$, is
constructed by generating Gaussian $A_{ij}$ and adopting 
the rank-ordering scheme
introduced recently\cite{Cieplak,Cieplak1} in the context of lattice
models. Sequence $R'$ is an off-lattice analog of the sequence $R$ discussed in
Refs.\cite{Cieplak,Cieplak1}. 
The principle here is to allocate the most
strongly attractive $A_{ij}$ 
to those pairs of monomers which form contacts in the
target compact lattice conformation. The true ground state of the 
sequence will necessarily be off-lattice but its shape will be
close to the lattice target -- it can be viewed as a somewhat distorted
lattice target conformation.

In Sec. III we describe the procedure adopted 
to determine the ground state. Basically,
we form self-avoiding walks (SAW) in continuum space around all of the
69 maximally compact lattice conformations as the starting configuration
of the sequence. We shall denote such starting SAW as CSAW to emphasize 
closeness to compact structures.
We then perform Monte Carlo
quenches to reach local energy minima. The rank-ordered allocation of
the couplings ensures that the lowest energy minima 
will be compact. For sequence $R'$ we have shown that the lowest
energy state is obtained from CSAW's which are near the target structure.
We demonstrate that the use of arbitrary SAW's, i.e., which are
not CSAW's, leads to local minima of higher energies.
For sequence $G$ both the SAW's and CSAW's starting
configurations easily lead to the native state which 
has the target compact shape. 
Our conclusion is that it is easier to find the native state for 
a good folder ($G$) compared to a bad folder ($R'$). We have found also
that only sufficiently large values of the spring constant $k$ may guarantee
the self-avoidance properties of the chain.

We characterize geometries of conformations in terms of a certain
distance away from a reference conformation.
In Sec. IV we 
determine the probability to stay in the basin and the folding temperature.
Both quantities are calculated by adopting a well defined characteristic
basin size as obtained by the shape distortion approach\cite{Li}.
In Sec. V we present results on the specific
heat and structural susceptibility for $R'$ and $G$
and demonstrate that $G$ is a good folder whereas $R'$ is a bad folder. 
The bad folding properties of $R'$ are in sharp contrast to what was found
for the lattice sequence $R$\cite{Cieplak,Cieplak1} and are due to the 
presence of many long-ranged couplings which impose conflicting constraints.
Sequence $G$ is a good folder because there are many fewer constraints 
to satisfy.

\section{Models of interactions}

Following IMP\cite{IMP}, we consider a self-interacting
heteropolymer in 2D described by the Hamiltonian given by
\begin{equation}
H\; \; = \; \; \sum_{i \neq j} \{ k (d_{i,j} - d_0)^2 \delta_{i,j+1}
+ 4 [ \frac{C}{d_{i,j}^{12}} - \frac{A_{ij}}{d_{i,j}^6} ] \} \; \; ,
\end{equation}
where $i$ and $j$ range from 1 to the number of beads,
$N$, which in our model is equal to 16. The distance between the beads,
$d_{i,j}$ is defined as $ |\vec{r}_{i}-\vec{r}_{j}|$, where
$\vec{r}_i$ denotes the position of bead $i$. The harmonic term
in the Hamiltonian with the spring constant $k$,
couples the adjacent beads along the chain.  The remaining terms 
represent the Lennard-Jones potential. 
In \cite{IMP} $A_{ij}$ is chosen as $A_{ij} = A_0 + \sqrt{\epsilon}\eta_{ij}$, 
where $A_0$ is constant
and $\eta_{ij}$'s are Gaussian variables with zero mean and unit variance;
$\epsilon$ controls the strength of the quenched disorder. 
The case of $\eta_{ij}=0$ and $A_0= C$ would correspond to a
homopolymer with the standard Lennard-Jones interaction used in 
simulations of liquids. We adopt the units in which $C$=1 and
consider $k$ to be either equal to 1 or to 25. We have found
that the first choice, which has been used by IMP\cite{IMP}, may
lead to local energy minima in which the polymer is self-intersecting.

\begin{figure}
\epsfxsize=3.2in
\centerline{\epsffile{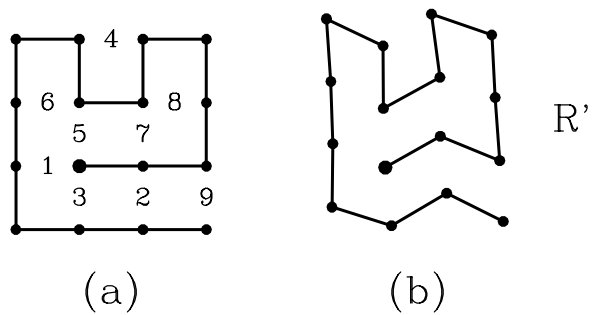}}
\caption{
(a) Assigment of nine strongest couplings $A_{ij}$ to native contacts
 for sequence
$R'$. The numbers indicate the relative strengths
of the contacts.
(b) The native conformation of sequence $R'$.}
\end{figure}

For $N$=16, there are 120 Lennard-Jones possible couplings between the monomers.
The basic choice for the values of $A_{ij}$ that we shall use here
is shown in Table I. Here, all of the $A_{ij}$'s are positive,
which corresponds to attraction. 
Table 1 also indicates a rank of a given attractive coupling if these are
rank ordered from the strongest to the weakest attraction.
In nine of the couplings, the attraction
is enhanced by making the corresponding $A_{ij}$ bigger than one.
These were chosen to coincide with the contacts present in the 
lattice target $R$' shown in Fig. 1(a).
The strongest attraction was assigned to be between
beads 1 and 12 and the relative strengths of other native attractions  are
indicated in the figure. The remaining 111 couplings are assigned $A_{ij}$
with values which are smaller than 1. Overall the 
mean value $A_0$ is equal to 0.784 and the dispersion to 0.205.

The parameter $d_0$ corresponding to the equilibrium distance in the
harmonic potential is chosen to be equal to 1.16, which is close to the
equilibrium position of the average Lennard-Jones potential, 
$(2C/A_0)^{1/6}$. The target lattice shapes are built on a lattice with this
lattice constant. 
For the Go-like sequence ($G$) we set $A_{ij}=1$ for native contact
and 0 for non-native ones. For this sequence we choose $d_0=2^{1/6}$.
The qualitative results do not depend, however, on the choice of $d_0$.

\section{Local minima and native states}

In order to find  spectra of the local energy minima
we use the Monte Carlo procedure
with local updating moves. 
Monomers are moved randomly within a circle (the radius of circle varies from
0.0025$d_0$ to 0.01$d_0$)
away from their previous positions. Random
quenching by starting from a SAW or a CSAW typically yields a local
energy minimum within of order $2^{18}$ steps. In the minimum,
the force acting on any of the monomers is at most of order $10^{-7}$
in units of the characteristic coupling (of order 1) divided
by $2C/A_0)^{1/6}$.
Simulated annealing runs of comparable number of steps yielded
similar results. The results 
reported on in this paper are based on the quenching procedure.

The starting conformations were obtained by placing monomers
within a circle of radius 0.3$d_0$ away from a SAW generated on the
lattice or away from a maximally compact conformation on the lattice
-- CSAW. For each of the 69 maximally conformations, 50 CSAW's
were generated and the results were compared with those
obtained based on 500 SAW's.


We observe that there is a substantial gap, of order 4,
between the minima obtained from CSAW's around the target and those
obtained  from all other starting configurations.


\begin{figure}
\epsfxsize=3.2in
\centerline{\epsffile{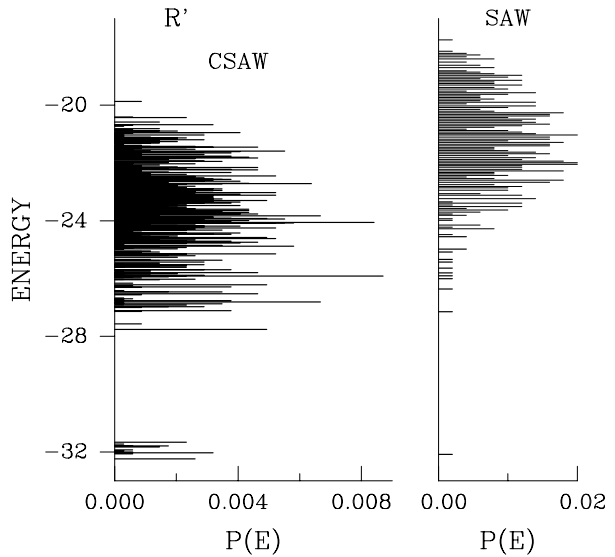}}
\caption{
Density of local minima for sequence $R'$ and $k=25$.
The left part corresponds
to those obtained from deformed compact cells as starting conformations
whereas the right part corresponds to the case when
starting conformations are arbitrary self-avoiding chains.
The energy gap is equal to 0.15.
(For $k=1$ above $E\approx -23.93$
the distance between the beads
may become bigger than the equilibrium distance of the Lennard-Jones potential
by 10\%. This happens in about 1.4\% of all conformations.)}
\end{figure}

As to the choice of the elastic constant $k$ we have found that
$k$=1 is not very
physical because the distances between consecutive monomers are not
kept sufficiently rigid. 
Furthermore, we have observed that
in states which are not local energy minima, the polymer conformations
may become self-intersecting. 
This phenomenon was also observed in Ref.\cite{Fukugita} for shorter chains.
Thus a stronger $k$ is needed and we focus on $k$=25.


The energy histograms for the $R'$ and $G$ sequences
for $k$=25 are shown in Figs. 2 and 3, respectively. 
In the case of $R'$, the SAW configurations happen to yield
results comparable to the CSAW  configurations, but the statistical
frequency of success in finding a low energy state favors the 
CSAW-based approach.
The maximal distance between monomers does not exceed $d_0$ by
more than 4\% in any local energy minimum, which 
demonstrates very good self-avoiding properties.

In the case of sequence $G$, about 25\% of CSAW's trajectories 
and 12\% of SAW's trajectories lead to the native conformation.
Thus it is easier to find the native state for $G$ than for $R'$.

The native state conformation of sequence $R'$ is shown in Fig 1(b).
The $x-y$ coordinates of the beads  of this conformation are listed in
Table II. They are used in the studies of kinetics and thermodynamics
properties. The native state of sequence $G$ has exactly the lattice shape
as shown in Fig. 1(a) because it is only the native contact energies here
that one needs to minimize. 


\begin{figure}
\epsfxsize=3.2in
\centerline{\epsffile{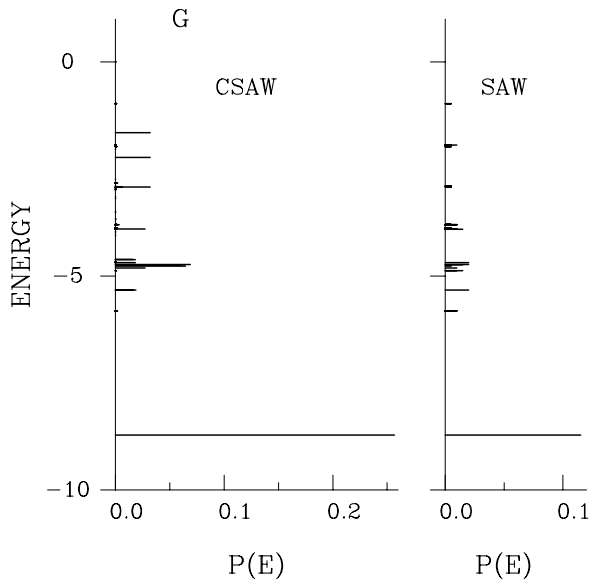}}
\caption{
The same as in Fig. 2 but for sequence $G$.
The gap is about 2.9.
}
\end{figure}

As we shall show later, several tasks, like the determination
of the folding temperature, are facilitated by introducing a
notion of a geometrical distance between two conformations. A convenient
definition of the distance $\delta_{ab}$ between two conformations
$a$ and $b$ is provided by 

\begin{equation}
\delta_{ab}^2 \; \; = \; \; $min$ \frac{1}{N} \sum_{i=1}^N 
| \vec{r}_i^a - \vec{r}_i^b |^2 \; \; ,
\end{equation}
where $\vec{r}_i^{a,b}$ denotes the position of monomer $i$ in 
conformation $a$($b$). The minimization is  performed over translations,
rotations and reflections. In practice, we put chain $a$ over chain $b$ by
overlapping the two centers of mass, and then we find the optimal rotation
of $b$ which minimizes $\delta_{ab}$. We pick the optimal angle from
1000 discrete values into which the 360$^o$ angle may be divided.

The distances between the native conformations 
and their lattice counter parts, as in Fig. 1,
are found to be equal to 0.296
and $\approx 0$ for sequence $R'$ and $G$ respectively.

\section{The folding temperature}

We now proceed to the equilibrium characterization of the sequences
introduced in this paper. The parameter that plays a primary role
in determining the folding characteristics is the folding
temperature $T_f$. In the case of lattice models, $T_f$ may 
be defined\cite{Dill} as the temperature at which the probability
of occupying the native state becomes equal to $\frac{1}{2}$.
For the  off-lattice model, however, the native state
has, strictly speaking, zero measure and one should deal with
the probability of occupying the native valley.

\begin{figure}
\epsfxsize=3.2in
\centerline{\epsffile{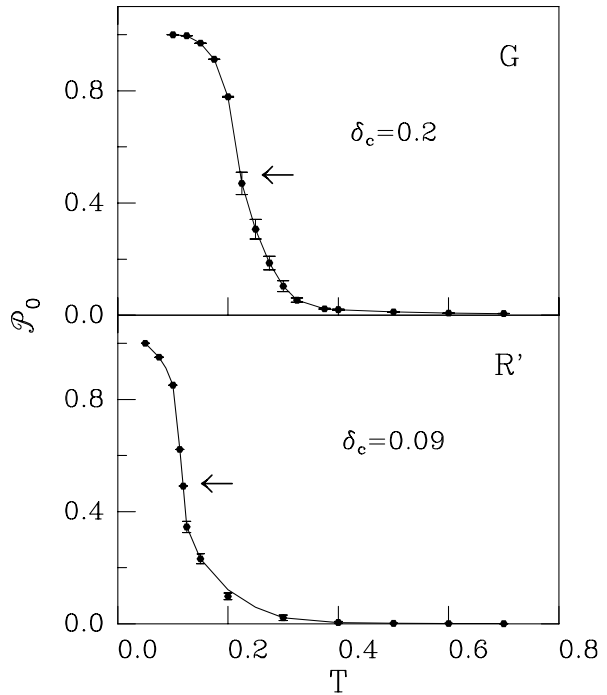}}
\caption{
The dependence of ${\cal P}_0$ on T for sequence $G$ and $R'$.
The arrow indicates the position of the
folding temperature which is equal to
$T_f=0.23$ for sequence $G$ and $T_f \approx 0.12$ for $R'$.
The results are
averaged over 10 - 40 Monte Carlo trajectories.}
\end{figure}

One may define the folding temperature $T_f$ through the following procedure. 
Suppose that $\delta$ is the distance to the 
native state and ${\cal P}(\delta)$ is the probability for a conformation
to be in this distance away from the native state.
Thus the probability to find the system in the immediate vicinity of
the native state is given by

\begin{equation}
{\cal P}_0 \; \; = \; \; \int_{0}^{\delta_{c}} {\cal P}(\delta) \, 
d\delta\; \; ,
\end{equation}
where $\delta_{c}$ is a cutoff distance. The folding temperature
$T_f$ is then defined as the temperature at which ${\cal P}_0=1/2$.

The size of the native basin $\delta_c$ was estimated by the shape distortion
approach\cite{Li}. In this approach one starts from the native state and
performs random displacements of invidual beads in the chain, through a
Monte Carlo routine. 
The distance to the native state is calculated by using Eq.(2) and the results
are averaged over many Monte carlo trajectories.
Below some critical temperature which may be interpreted
as a folding temperature $T_f$, the distance to the native state gets 
saturated at sufficiently long time scales. The satuation value of this distance
at the critical temperature can serve as the size of the native 
basin $\delta_c$.
It is found that $\delta_c \approx 0.2$ and $\delta_c \approx 0.09$ for
$G$ and $R'$, respectively\cite{Li}. The corresponding values of the
folding temperatures obtained in this way are equal to $T_f\approx 0.19$ 
and $T_f\approx 0.09$
for $G$ and $R'$\cite{Li}.

\begin{figure}
\epsfxsize=3.2in
\centerline{\epsffile{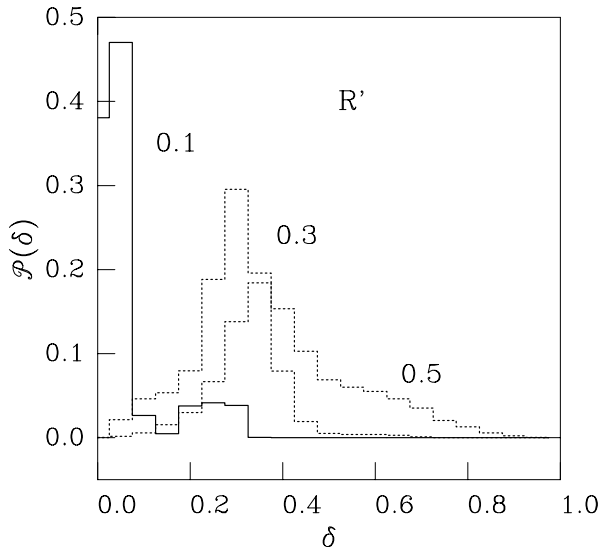}}
\caption{
The probability distribution function ${\cal P}(\delta)$  to find
a state which is in distance $\delta$ away from the native state
for sequence $R'$. The values
of $T$ are indicated next to the curves.
}
\end{figure}

Figure 4 shows the probability to be in the native valley versus
temperature  for $G$ and $R'$. ${\cal P}_0$ has been
determined by estimating the probability to stay in the native valley
by starting in the native state and monitoring the system for 5$\times 10^6$
Monte Carlo steps (the distance updating is of order 0.01).
Doubling the length of the run does not affect the results in any
visible manner. 
For sequence $G$ and $R'$, we have found $T_f\approx 0.23$ and 
$T_f\approx 0.12$. These values of $T_f$ are close to those found by the 
shape distortion method\cite{Li}.

Figure 5 shows the distribution ${\cal P}(\delta)$ at different 
temperatures for  sequence $R'$.
As the temperature increases the maximum
becomes wider and moves toward larger distances to the native state $\delta$.

We now turn to the temperature dependence of the folding time, $t_{fold}$.
At high temperatures, reaching the native state takes long due to
entropic effects. At low temperatures, on the other hand, glassy phenomena
may set in and make the folding process extremely slow. 
Thus $t_{fold}$ plotted against $T$ typically shows a minimum at a certain
temperature $T_{min}$.
The idea of the existence
of the glassy phase in proteins has originated in the Bryngelson and Wolynes
studies of the random energy model \cite{Bryngelson,Bryngelson1} and was
subsequently tested in numerical simulations 
of lattice models by Socci and Onuchic \cite{Socci}.
Notice that $T_f$ is a characteristic temperature that relates to
equilibrium whereas $T_{min}$ is a characteristic temperature that relates
to dynamical properties. Even though at $T= T_{min}$ folding is the fastest
this temperature also marks the onset of the glassy effects
because they become stronger and stronger on departing from
$T_{min}$ towards lower temperatures.
Thus if $T_f$ is significantly less than $T_{min}$ a sequences is bad folder.
If $T_f$ is comparable to $T_{min}$, or preferably larger than $T_{min}$,
then the sequence is a good folder\cite{Socci}.
It should be noted that
as a characteristic temperature that relates to dynamics
one often uses the glass transition temperature
$T_g$ \cite{Dill,Socci}.
$T_g$ is, however, depends on
a cutoff value of the time used in calculations. Our preference here is to use 
$T_{min}$, instead of $T_g$, not only because its definition is unique and
independent of the value of the cutoff but 
also because the two quantities contain the same the physics: good folders
are sequences in which glassy effect are not important around the folding
temperature.

In order to obtain $t_{fold}$ we start from random 
configurations and 
evolve them through  the Monte Carlo process
until the native state is reached. 
$t_{fold}$ is defined as
the median number of MC steps after which the system reaches
 the basin of the native state for the
first time. The cutoff value of MC steps is taken to be 20 millions. 
The number of starting configurations varies from 20 to 40 depending on $T$.
Our results are presented in Fig. 6 for sequence $R'$ and $G$.
We have $T_{min}=0.4\pm 0.1$ and $T_{min}=0.15\pm 0.05$ for $R'$ and $G$
respectively. Since $T_f=0.12$ and $T_f=0.23$ for these 
sequences one can see that $R'$ is bad folder whereas $G$ is a good one.
We emphasize that on-lattice Go and rank-ordering schemes lead to comparable 
ratios of $T_f/T_{min}$.

\begin{figure}
\epsfxsize=3.2in
\centerline{\epsffile{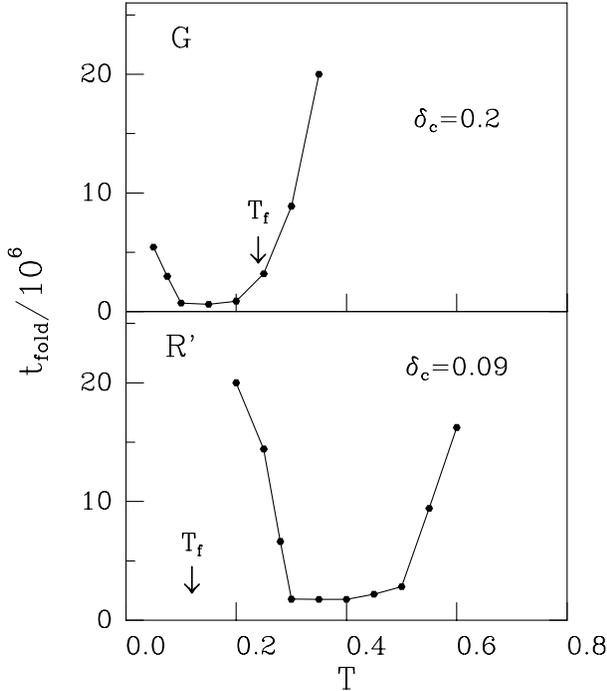}}
\caption{
The temperature dependence of folding time for sequence $G$ and
$R'$.
The results are obtained using 20 - 50 starting configurations.}
\end{figure}

\section{Thermodynamic properties}

We now proceed to a further characterization of the sequences by considering
the thermodynamic quantities such
as the specific heat, $C$, and the structural susceptibility, $\chi$, 
\cite{Camacho}. It has been suggested\cite{Camacho,Camacho1,Thirumalai} 
that a small temperature
difference between the maximum in $C$ and the
maximum in $\chi$ when plotted against $T$ indicates good folding properties.
Thus studies of these two quantities may serve as a substitute
for the information about the kinetics of folding.

In the case of of an off-lattice model the departures of 
the sequence geometry from its native conformation may be desribed through the 
structural overlap function\cite{Klimov} as:
\begin{equation}
\chi_s \; \; = \; \; 1 - \frac{1}{N^2-3N+1}
\sum_{i\ne j,j\pm1} \; \Theta (\delta_c-|r_{ij}-r^N_{ij}|) \; ,
\end{equation}
where $r_{ij}$ is the distance between the beads $i$ and $j$ for a given
conformation, $r_{ij}^N$ is the corresponding distance in the native 
conformation, and $\Theta (x)$ is the Heavyside function. Here $\delta_c$
denotes the size of basin as defined above in the previous Section. 
If $|r_{ij}-r_{ij}^N|\le \delta_c$
the beads $i$ and $j$ are assumed to form a contact within the native valley.
The susceptibility-like parameter $\chi$ is  defined as the thermal fluctuation
of $\chi_s$:
\begin{equation}
\chi(T) \; \; = \; \; < \chi_s^2(T) > - < \chi_s(T) >^2 \; ,
\end{equation}
where the angular brackets indicate a thermodynamic average.

\begin{figure}
\epsfxsize=3.2in
\centerline{\epsffile{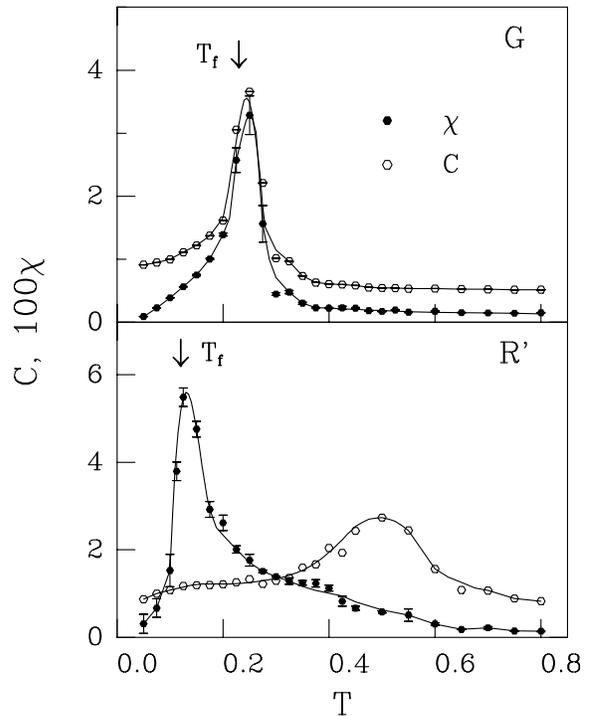}}
\caption{
The temperature dependence of $\chi$ and $C$ for $G$ and $R'$. The starting
configurations are the native state. The results are averaged over 10 MC
trajectories. The arrows indicate the position of
$T_f$ found from the condition that the probability to stay
in the native state is equal to 1/2.}
\end{figure}

The specific heat is defined in the usual way, i.e.
by the energy fluctuations:
\begin{equation}
C \; \; = \; \; \frac{<E^2> - <E>^2}{T^2} \; .
\end{equation}
A peak in $C$ may be interpreted as corresponding to the onset
of slow kinetics and the peak in $\chi$ as corresponding to the
folding temperature.

We calculate the thermodynamic quantities using the native state as 
the starting configurations. The results are averaged over many
MC trajectories. The equilibration is checked by monitoring the stability
of the data against at least three-times longer runs. 
We have used  5$\times 10^6$ MC steps  as in the studies of ${\cal P}_0$
(the first 2.5$\times 10^6$ steps are not taken into account when averaging).
This number of MCS appears to be enough to equilibrate the system in 
the temperature interval we use. We have also found that it is harder to
reach the equilibrium using SAW's as starting configurations.

The temperature dependence of $\chi$ and $C$ for sequence
$G$ and $R'$ is shown in Fig. 7.
The peak in $\chi$  almost coincides
with $T_f$. For $G$ maxima of $\chi$ and $C$ are located at the same position
suggesting that sequence $G$ is a good folder\cite{Camacho}.
This results agrees with that obtained in the previous
section by studying kinetics of folding.


For sequence $R'$, the peak in $C$ is found to be broad and it is located 
at a substantially higher $T$ than the peak in $\chi$ confirming
bad folding properties of $R'$.
So studies of thermodynamics properties and folding kinetics
lead to the same conclusion about folding in our model.

\section{Conclusions}

In this work, we have constructed and studied two off-lattice Lennard-Jones
models in two dimensions.
The long-range nature of interactions between amino-acids in off-lattice models
may lead to too much frustration, as in the case of sequence $R'$, and thus 
to bad foldicities.
Restricting the number of interactions only to the native contacts, 
as in sequence $G$, 
may reduce the frustration and bring about good folding properties.
We have demonstrated this by studying both thermodynamic and dynamical 
properties of the two sequences.
The values of the folding temperature found from the probability to get out 
from the native state roughly agree with those obtained by the shape distortion
approach.
We have studied the kinetic of folding with a simplified MC dynamics
whereas previous works have been focused on equilibrium aspects.
Similar to the on-lattice models, both approaches give the same
information about folding properties.

We thank J. R. Banavar for discussions.
This work was supported by Komitet Badan Naukowych (Poland; Grant number\
2P03B-2513).

\vspace{1cm}


\twocolumn[\hsize\textwidth\columnwidth\hsize\csname @twocolumnfalse\endcsname
\begin{table}
\begin{center}
  \vskip 0.5cm
\begin{tabular}{|l|l|l|l||l|l|l|l||l|l|l|l||l|l|l|l|} \hline
   $i$&   $j$&$\; \; \; \; A_{ij}$&\hspace{0.3cm} rank\hspace{0.3cm}&  $i$&  $j$ & $\; \; \; \; A_{ij}$&\hspace{0.3cm} rank\hspace{0.3cm}&   $i$&  $j$& $ \; \; \; \; A_{ij}$&\hspace{0.3cm}  rank\hspace{0.3cm}&   $i$&  $j$& $ \; \; \; \; A_{ij}$&\hspace{0.3cm}  rank\hspace{0.3cm}\\ \hline
   1&   2&  0.946554&\hspace{0.3cm}  19\hspace{0.3cm}&   3&   5&  0.581629&\hspace{0.3cm} 105\hspace{0.3cm}&   5&  12&  0.667698&\hspace{0.3cm}  79\hspace{0.3cm}&   8&  15&  0.758490&\hspace{0.3cm}  57\hspace{0.3cm}\\
   1&   3&  0.858677&\hspace{0.3cm}  37\hspace{0.3cm}&   3&   6&  0.598687&\hspace{0.3cm} 101\hspace{0.3cm}&   5&  13&  0.740622&\hspace{0.3cm}  62\hspace{0.3cm}&   8&  16&  0.718888&\hspace{0.3cm}  69\hspace{0.3cm}\\ 
   1&   4&  0.810776&\hspace{0.3cm}  47\hspace{0.3cm}&   3&   7&  0.513536&\hspace{0.3cm} 118\hspace{0.3cm}&   5&  14&  0.661444&\hspace{0.3cm}  83\hspace{0.3cm}&   9&  10&  0.913366&\hspace{0.3cm}  28\hspace{0.3cm}\\
   1&   5&  0.949365&\hspace{0.3cm}  17\hspace{0.3cm}&   3&   8&  0.778139&\hspace{0.3cm}  50\hspace{0.3cm}&   5&  15&  0.861819&\hspace{0.3cm}  36\hspace{0.3cm}&   9&  11&  0.741376&\hspace{0.3cm}  61\hspace{0.3cm}\\
   1&   6&  0.514208&\hspace{0.3cm} 117\hspace{0.3cm}&   3&   9&  0.746676&\hspace{0.3cm}  59\hspace{0.3cm}&   5&  16&  0.744776&\hspace{0.3cm}  60\hspace{0.3cm}&   9&  12&  0.664109&\hspace{0.3cm}  82\hspace{0.3cm}\\ 
   1&   7&  0.644279&\hspace{0.3cm}  91\hspace{0.3cm}&   3&  10&  0.666580&\hspace{0.3cm}  80\hspace{0.3cm}&   6&   7&  0.767029&\hspace{0.3cm}  54\hspace{0.3cm}&   9&  13&  0.968419&\hspace{0.3cm}  11\hspace{0.3cm}\\
   1&   8&  1.337928&\hspace{0.3cm}   5\hspace{0.3cm}&   3&  11&  0.835338&\hspace{0.3cm}  41\hspace{0.3cm}&   6&   8&  0.638391&\hspace{0.3cm}  93\hspace{0.3cm}&   9&  14&  0.822724&\hspace{0.3cm}  45\hspace{0.3cm}\\
   1&   9&  0.725085&\hspace{0.3cm}  65\hspace{0.3cm}&   3&  12&  0.722943&\hspace{0.3cm}  67\hspace{0.3cm}&   6&   9&  1.375767&\hspace{0.3cm}   4\hspace{0.3cm}&   9&  15&  0.969623&\hspace{0.3cm}  10\hspace{0.3cm}\\
   1&  10&  0.802427&\hspace{0.3cm}  48\hspace{0.3cm}&   3&  13&  0.518711&\hspace{0.3cm} 116\hspace{0.3cm}&   6&  10&  0.676479&\hspace{0.3cm}  76\hspace{0.3cm}&   9&  16&  0.850939&\hspace{0.3cm}  39\hspace{0.3cm}\\  
   1&  11&  0.763514&\hspace{0.3cm}  55\hspace{0.3cm}&   3&  14&  0.660732&\hspace{0.3cm}  85\hspace{0.3cm}&   6&  11&  0.567934&\hspace{0.3cm} 109\hspace{0.3cm}&  10&  11&  0.935812&\hspace{0.3cm}  22\hspace{0.3cm}\\
   1&  12&  1.498411&\hspace{0.3cm}   1\hspace{0.3cm}&   3&  15&  0.852925&\hspace{0.3cm}  38\hspace{0.3cm}&   6&  12&  0.571524&\hspace{0.3cm} 108\hspace{0.3cm}&  10&  12&  0.951547&\hspace{0.3cm}  16\hspace{0.3cm}\\
   1&  13&  0.899355&\hspace{0.3cm}  30\hspace{0.3cm}&   3&  16&  1.050544&\hspace{0.3cm}   9\hspace{0.3cm}&   6&  13&  0.658820&\hspace{0.3cm}  86\hspace{0.3cm}&  10&  13&  0.521468&\hspace{0.3cm} 114\hspace{0.3cm}\\ 
   1&  14&  1.396376&\hspace{0.3cm}   3\hspace{0.3cm}&   4&   5&  0.878341&\hspace{0.3cm}  33\hspace{0.3cm}&   6&  14&  0.512926&\hspace{0.3cm} 119\hspace{0.3cm}&  10&  14&  0.930277&\hspace{0.3cm}  25\hspace{0.3cm}\\
   1&  15&  0.723510&\hspace{0.3cm}  66\hspace{0.3cm}&   4&   6&  0.763190&\hspace{0.3cm}  56\hspace{0.3cm}&   6&  15&  0.612393&\hspace{0.3cm}  99\hspace{0.3cm}&  10&  15&  0.889378&\hspace{0.3cm}  31\hspace{0.3cm}\\
   1&  16&  0.655780&\hspace{0.3cm}  87\hspace{0.3cm}&   4&   7&  1.269149&\hspace{0.3cm}   8\hspace{0.3cm}&   6&  16&  0.628334&\hspace{0.3cm}  95\hspace{0.3cm}&  10&  16&  0.590144&\hspace{0.3cm} 104\hspace{0.3cm}\\ 
   2&   3&  0.720592&\hspace{0.3cm}  68\hspace{0.3cm}&   4&   8&  0.577229&\hspace{0.3cm} 106\hspace{0.3cm}&   7&   8&  0.670401&\hspace{0.3cm}  78\hspace{0.3cm}&  11&  12&  0.932459&\hspace{0.3cm}  23\hspace{0.3cm}\\
   2&   4&  0.541154&\hspace{0.3cm} 112\hspace{0.3cm}&   4&   9&  0.614099&\hspace{0.3cm}  96\hspace{0.3cm}&   7&   9&  0.647826&\hspace{0.3cm}  89\hspace{0.3cm}&  11&  13&  0.528390&\hspace{0.3cm} 113\hspace{0.3cm}\\
   2&   5&  0.770339&\hspace{0.3cm}  53\hspace{0.3cm}&   4&  10&  0.918839&\hspace{0.3cm}  27\hspace{0.3cm}&   7&  10&  0.642560&\hspace{0.3cm}  92\hspace{0.3cm}&  11&  14&  0.921718&\hspace{0.3cm}  26\hspace{0.3cm}\\
   2&   6&  0.938498&\hspace{0.3cm}  21\hspace{0.3cm}&   4&  11&  0.597736&\hspace{0.3cm} 102\hspace{0.3cm}&   7&  11&  0.965115&\hspace{0.3cm}  12\hspace{0.3cm}&  11&  15&  0.770826&\hspace{0.3cm}  52\hspace{0.3cm}\\
   2&   7&  1.291634&\hspace{0.3cm}   7\hspace{0.3cm}&   4&  12&  0.571611&\hspace{0.3cm} 107\hspace{0.3cm}&   7&  12&  0.834125&\hspace{0.3cm}  42\hspace{0.3cm}&  11&  16&  0.955652&\hspace{0.3cm}  14\hspace{0.3cm}\\ 
   2&   8&  0.696944&\hspace{0.3cm}  72\hspace{0.3cm}&   4&  13&  0.561490&\hspace{0.3cm} 110\hspace{0.3cm}&   7&  13&  0.753538&\hspace{0.3cm}  58\hspace{0.3cm}&  12&  13&  0.629158&\hspace{0.3cm}  94\hspace{0.3cm}\\
   2&   9&  0.603317&\hspace{0.3cm} 100\hspace{0.3cm}&   4&  14&  0.725428&\hspace{0.3cm}  64\hspace{0.3cm}&   7&  14&  0.561397&\hspace{0.3cm} 111\hspace{0.3cm}&  12&  14&  0.884585&\hspace{0.3cm}  32\hspace{0.3cm}\\
   2&  10&  0.832387&\hspace{0.3cm}  43\hspace{0.3cm}&   4&  15&  0.954825&\hspace{0.3cm}  15\hspace{0.3cm}&   7&  15&  0.712665&\hspace{0.3cm}  70\hspace{0.3cm}&  12&  15&  0.949139&\hspace{0.3cm}  18\hspace{0.3cm}\\
   2&  11&  0.672241&\hspace{0.3cm}  77\hspace{0.3cm}&   4&  16&  0.511493&\hspace{0.3cm} 120\hspace{0.3cm}&   7&  16&  0.864981&\hspace{0.3cm}  35\hspace{0.3cm}&  12&  16&  0.932023&\hspace{0.3cm}  24\hspace{0.3cm}\\ 
   2&  12&  0.821738&\hspace{0.3cm}  46\hspace{0.3cm}&   5&   6&  0.687186&\hspace{0.3cm}  74\hspace{0.3cm}&   8&   9&  0.943903&\hspace{0.3cm}  20\hspace{0.3cm}&  13&  14&  0.848886&\hspace{0.3cm}  40\hspace{0.3cm}\\
   2&  13&  0.644946&\hspace{0.3cm}  90\hspace{0.3cm}&   5&   7&  0.665046&\hspace{0.3cm}  81\hspace{0.3cm}&   8&  10&  0.660797&\hspace{0.3cm}  84\hspace{0.3cm}&  13&  15&  0.613872&\hspace{0.3cm}  97\hspace{0.3cm}\\
   2&  14&  0.689698&\hspace{0.3cm}  73\hspace{0.3cm}&   5&   8&  0.686304&\hspace{0.3cm}  75\hspace{0.3cm}&   8&  11&  1.301709&\hspace{0.3cm}   6\hspace{0.3cm}&  13&  16&  0.734204&\hspace{0.3cm}  63\hspace{0.3cm}\\ 
   2&  15&  1.409250&\hspace{0.3cm}   2\hspace{0.3cm}&   5&   9&  0.771102&\hspace{0.3cm}  51\hspace{0.3cm}&   8&  12&  0.903380&\hspace{0.3cm}  29\hspace{0.3cm}&  14&  15&  0.519439&\hspace{0.3cm} 115\hspace{0.3cm}\\
   2&  16&  0.655129&\hspace{0.3cm}  88\hspace{0.3cm}&   5&  10&  0.595747&\hspace{0.3cm} 103\hspace{0.3cm}&   8&  13&  0.612414&\hspace{0.3cm}  98\hspace{0.3cm}&  14&  16&  0.963250&\hspace{0.3cm}  13\hspace{0.3cm}\\ 
   3&   4&  0.781090&\hspace{0.3cm}  49\hspace{0.3cm}&   5&  11&  0.707574&\hspace{0.3cm}  71\hspace{0.3cm}&   8&  14&  0.825123&\hspace{0.3cm}  44\hspace{0.3cm}&  15&  16&  0.866528&\hspace{0.3cm}  34\hspace{0.3cm}\\ \hline
\end{tabular}
\vskip 0.5cm
\caption{Values of the $A_{ij}$ for sequence $R'$.
The rank of a
coupling is indicated to the right of its value.}
\label{table1}
\end{center}
\end{table}
]

\twocolumn[\hsize\textwidth\columnwidth\hsize\csname @twocolumnfalse\endcsname
\begin{table}
\begin{center}
  \vskip 0.5cm
\begin{tabular}{|l|l|l||l|l|r|} \hline
\hspace{0.25cm} monomer \hspace{1.75cm}& $\; \; \; \; \; x_N$&$\; \; \; \; y_N  \;$ & \hspace{0.25cm} monomer \hspace{1.75cm}& $\; \; \; \; \; x_N$&$y_N \; \; \; \;
$\\ \hline

\hspace{1cm}   1 \hspace{1cm}&$\; \; \; \; \;$ 0 &0 $\; \; \; \; \;$ & \hspace{1cm}   9 \hspace{1cm}&  -0.037572&   2.225111\\
\hspace{1cm}   2 \hspace{1cm}&   1.005509&   0.573347 & \hspace{1cm}  10 \hspace{1cm}&  -1.073118&   2.718455\\
\hspace{1cm}   3 \hspace{1cm}&   2.091357&   0.126744 & \hspace{1cm}  11 \hspace{1cm}&  -0.992672&   1.573907\\
\hspace{1cm}   4 \hspace{1cm}&   2.009697&   1.279681 & \hspace{1cm}  12 \hspace{1cm}&  -0.956188&   0.436143\\
\hspace{1cm}   5 \hspace{1cm}&   1.946778&   2.424499 & \hspace{1cm}  13 \hspace{1cm}&  -0.972265&  -0.723659\\
\hspace{1cm}   6 \hspace{1cm}&   0.845883&   2.806329 & \hspace{1cm}  14 \hspace{1cm}&   0.116648&  -1.062284\\
\hspace{1cm}   7 \hspace{1cm}&   0.997773&   1.647486 & \hspace{1cm}  15 \hspace{1cm}&   1.123476&  -0.470181\\
\hspace{1cm}   8 \hspace{1cm}&  -0.035144&   1.079819 & \hspace{1cm}  16 \hspace{1cm}&   2.154153&  -0.984088\\ \hline
\end{tabular}
\vskip 0.5cm
\caption{The $x-y$ coordinates of the native conformation
for $R'$ and $k=25$.}
\label{table2}
\end{center}
\end{table}
]

\end{document}